\documentclass[a4paper]{article}
\usepackage{amsmath,amssymb,amsfonts,amsthm,amscd}
\usepackage[titletoc,title]{appendix}
\usepackage[bookmarks=true]{hyperref}
\usepackage{color}
\numberwithin{equation}{section}

\newcommand{\bea}{\begin{eqnarray}}
\newcommand{\eea}{\end{eqnarray}}
\newcommand{\be}{\begin{equation}}
\newcommand{\ee}{\end{equation}}
\newcommand{\nn}{\nonumber \\}

\begin{document}
\begin{titlepage}

\vfill \vfill \vfill
\begin{center}
{\bf\Large Couplings of ${\cal N}=4$, $d=1$ mirror supermultiplets}
\end{center}
\vspace{1.5cm}

\begin{center}
{\large\bf Evgeny Ivanov${\,}^{a)\,b)}$, Stepan Sidorov${\,}^{a)}$}
\end{center}
\vspace{0.4cm}

\centerline{${\,}^{a)}$ \it Bogoliubov Laboratory of Theoretical Physics, JINR, 141980 Dubna, Moscow Region, Russia}
\vspace{0.2cm}
\centerline{${\,}^{b)}$ \it  Moscow Institute of Physics and Technology,
141700 Dolgoprudny, Moscow region, Russia}
\vspace{0.3cm}

\centerline{\tt eivanov@theor.jinr.ru, sidorovstepan88@gmail.com}
\vspace{0.2cm}

\vspace{2cm}

\par

{\abstract
We construct models of coupled semi-dynamical (spin) and dynamical mirror multiplets of ${\cal N}=4$ supersymmetric mechanics in  $d=1$ harmonic superspace.
Specifically, we consider a semi-dynamical mirror multiplet ${\bf (3,4,1)}$ coupled to dynamical mirror multiplets ${\bf (1, 4, 3)}$ and ${\bf (2, 4, 2)}$.
Coupling of the multiplets ${\bf (3, 4, 1)}$ and ${\bf (1, 4, 3)}$ yields a mirror counterpart of
the earlier constructed model implying the Nahm equations for the spin variables with the bosonic component of the multiplet ${\bf (1, 4, 3)}$ as an evolution parameter.
We also couple the mirror multiplet ${\bf (2,4,2)}$ to the mirror semi-dynamical multiplet ${\bf (3, 4, 1)}$ using chiral ${\cal N}=4$ superspace.
The models constructed admit a generalization to the SU$(2|1)$ deformation of ${\cal N}=4$, $d=1$ Poincar\'e supersymmetry.
\noindent }

\vfill{}


\end{titlepage}
\section{Introduction}

Diverse models of the supersymmetric quantum mechanics (SQM) as an extreme (one-dimensional) supersymmetric theory provide a good laboratory for studying more ambitious
higher-dimensional supersymmetric theories, such as super Yang-Mills and supergravity theories, the higher-spin theories, etc (see, e.g., \cite{Rev} for a review).
The simplest extended SQM models are associated with ${\cal N}=4, d=1$ supersymmetry. One of the surprising features of this supersymmetry is the existence
of two different types of ${\cal N}=4$ supermultiplets which are ``mirror'' (or ``twisted'') with respect to each other.

The origin of such a doubling is as follows. The ${\cal N}=4, d=1$ Poincar\'e superalgebra reads:
\bea
    \left\lbrace Q^{i}_{\beta}, Q_{j}^{\alpha}\right\rbrace = 2\delta^i_j\delta^\alpha_\beta H,\qquad  \left[H, Q^{i}_{\beta}\right] = 0, \label{N4}
\eea
where $H$ is the Hamiltonian which in the superfield (or component) Lagrangian setting is realized as a time-derivative. Four supercharges $Q^{i}_{\beta}$
carry the indices of the fundamental representation of the corresponding automorphism
${\rm SU}(2)_{\rm L} \times {\rm SU}(2)_{\rm R}$ group ($i = 1,2$ and $\alpha = 1,2$).
Their permutation as $i, j\leftrightarrow \alpha, \beta$ has no impact on the algebra \eqref{N4}. As a result,
${\cal N}=4$, $d=1$ supersymmetry possesses two wide classes of the supermultiplets which differ just by interchanging of these two independent ${\rm SU}(2)$
factors of the total automorphism group. The mutual interchange of these two ${\rm SU}(2)$ groups
switches ordinary multiplets into mirror ones and {\it vice versa}. When limiting only to one type of such multiplets and considering their various invariant actions
and interactions, no actual difference from another type can be observed: indeed, all quantities associated with  the alternative choice can be reproduced from the initial
choice just by substituting the ${\rm SU}(2)_{\rm L}$ group indices altogether by the appropriate ${\rm SU}(2)_{\rm R}$ ones. The difference between the two varieties
of the multiplets manifests itself only when considering both types of them {\it simultaneously} \footnote{A similar phenomenon takes place for the standard and twisted
chiral superfields in $2D$ supersymmetry \cite{GaHuRo}.}.

For a fixed choice of ${\rm SU}(2)$ (${\rm SU}(2)_{\rm L} $ in what follows) plethora of relevant multiplets
was studied in many papers, using the appropriate ${\cal N}=4, d=1$ superspace approaches in which just this ${\rm SU}(2)$  invariance is manifest \footnote{The basic $d=1$ superspace technicalities are collected in Appendix \ref{AppA}.}.
We will refer to these multiplets as the ``ordinary'' ones. The best arena for dealing with such multiplets and constructing their interactions is provided
by ${\cal N}=4, d=1$ harmonic superspace \cite{HSS} involving the harmonic variables which parametrize the coset ${\rm SU}(2)_{\rm L}/{\rm U}(1)_{\rm L} $. The second ${\rm SU}(2)_{\rm R} $
symmetry is realized as a kind of hidden symmetry. On the other hand, in order to put the description of both types of ${\cal N}=4$ multiplets on equal footing,
the formalism of  ``bi-harmonic superspace'' was worked out in \cite{BHSS}, with both automorphism ${\rm SU}(2)$ factors being ``harmonized''. However, dealing with the two sets
of harmonic variables sometimes bears technical complications. So it would be advantageous to have a description of the mirror multiplets within the same superspace setting
as the more accustomed ``ordinary'' ${\cal N}=4$ multiplets. The present note is devoted to such an alternative description of mirror multiplets and demonstrating that various
interactions between them basically lead to the same component results as those for the ordinary multiplets, modulo the
interchange of the ${\rm SU}(2)_{\rm L}$ and ${\rm SU}(2)_{\rm R}$ automorphism groups mentioned above. We focus on couplings of the dynamical mirror multiplets
${\bf (1, 4, 3)}$ and ${\bf (2, 4, 2)}$ to the mirror multiplet ${\bf (3, 4, 1)}$ considered as semi-dynamical (or as a ``spin multiplet''),  because the chiral multiplet ${\bf (2, 4, 2)}$
was not considered before in such a context. Another reason is that just this kind of couplings admits a rather direct generalization to the case of deformed ${\cal N}=4, d=1$
supersymmetry associated with the supergroup ${\rm SU}(2|1)$ \cite{I,II,III}. We explicitly present the basic relations
of the ${\rm SU}(2|1)$ deformed mirror system ${\bf (1, 4, 3)} - {\bf (3, 4, 1)}$.



\section{Mirror multiplets}
We proceed from the standard ${\cal N}=4$, $d=1$ superspace and its simplest harmonic extension described in Appendix \ref{AppA}.

An important observation exploited in what follows is that all the standard mirror multiplets with four fermionic physical fields and linear ${\cal N}=4$ supersymmetry transformation
laws are described by the superfields $M$ which carry
no external SU$(2)_{\rm L}$ indices (but admit those of SU$(2)_{\rm R}$) and satisfy the universal common constraint
\bea
    D^{(i}_{\gamma}D^{j)\gamma}M=0.\label{mirror}
\eea
In the harmonic ${\cal N}=4$, $d=1$ superspace approach these superfields are neutral (with respect to the harmonic ${\rm U}(1)_{\rm L}$ charge)
and can be defined by the following equivalent constraints:
\bea
    D^{++}M=0,\qquad D^0M=0,\qquad D^{+}_{\gamma}D^{+\gamma}M=0.
\eea
The specificity of one or another mirror multiplet manifests itself in the extra constraints one needs to impose on $M$.
Below we list all ${\cal N}=4$ superfield constraints of this kind yielding the complete set of the linear mirror multiplets.
\vspace{0.2cm}

\begin{itemize}
\item[]{\bf Mirror multiplet ${\bf (1, 4, 3)}$.}
The mirror multiplet ${\bf (1, 4, 3)}$ is described by a real superfield $X$ satisfying \cite{DelIva2007}
\bea
    D^{(i}_{\alpha}D^{j)\alpha}X=0 \quad\Longleftrightarrow\quad D^{+}_{\alpha}D^{+\alpha}X=0,\qquad D^{++}X=0.\label{143}
\eea
So in this simplest case no any extra constraints are needed.
\vspace{0.2cm}

\item[]{\bf Mirror multiplet ${\bf (2, 4, 2)}$ or chiral multiplet.}
The mirror multiplet ${\bf (2, 4, 2)}$ is described by the standard complex chiral ${\cal N}=4, d=1$ superfield:
\bea
    \bar{D}^{i}Z=0, \; \bar{D}^{i} := D^{i\, \alpha=2} = D^i_{\alpha=1}\,,  \quad\Longleftrightarrow\quad D^{+\,\alpha=2}Z=0,\qquad  D^{++}Z = 0.\label{chiral}
\eea
Thus, in the universal description by a superfield $M$, it is natural to interpret  the standard chiral ${\cal N}=4$, $d=1$ multiplet as belonging to the mirror type,
while the twisted chiral multiplet studied in \cite{DelIva2007, FedSmi2015} should be reckoned to the set of ``ordinary'' multiplets.
\vspace{0.2cm}

\item[]{\bf Mirror multiplet ${\bf (3, 4, 1)}$.}
The mirror multiplet ${\bf (3, 4, 1)}$ is described by a triplet superfield
$V^{\alpha\beta}$ ($V^{\alpha\beta}=V^{\beta\alpha}$, $\overline{\left(V^{\alpha\beta}\right)}=-\,V_{\alpha\beta}$) satisfying
\bea
    D^{i(\alpha}V^{\beta\gamma)}=0 \quad\Longleftrightarrow\quad
    D^{+(\alpha}V^{\beta\gamma)}=0, \qquad D^{++}V^{\alpha\beta}=0.\label{341}
\eea
\vspace{0.2cm}

\item[]{\bf Mirror multiplet ${\bf (4, 4, 0)}$.}
The mirror ${\bf (4, 4, 0)}$ multiplet is described by a quartet superfield $Y^{\alpha A}$ ($A=1,2$) that satisfies the constraints
\bea
    D^{i(\alpha}Y^{\beta) A} =0,\qquad \overline{\left(Y^{\alpha A}\right)}=Y_{\alpha A}\,.\label{440}
\eea
Their equivalent  harmonic superspace form is
\bea
    D^{+(\alpha}Y^{\beta) A}=0,\qquad
    D^{++}Y^{\alpha A}=0.\label{440HSS}
\eea
\vspace{0.2cm}

\item[]{\bf Mirror multiplet ${\bf (0, 4, 4)}$.} In contrast to the multiplet ${\bf (4, 4, 0)}$, the mirror multiplet ${\bf (0, 4, 4)}$
is described by a fermionic superfield $\Psi^{\alpha A}$ ($A=1,2$) \cite{IvanSigma}:
\bea
    D^{i(\alpha}\Psi^{\beta) A} =0,\qquad \overline{\left(\Psi^{\alpha A}\right)}=\Psi_{\alpha A}\quad\Rightarrow\quad
    D^{+(\alpha}\Psi^{\beta) A}=0,\qquad
    D^{++}\Psi^{\alpha A}=0.\label{044}
\eea
All its  bosonic components are auxiliary fields.
\end{itemize}

The component solutions of the constraints for the multiplets ${\bf (1, 4, 3)}$, ${\bf (2, 4, 2)}$, and ${\bf (3, 4, 1)}$ are given by eqs. \eqref{X143}, \eqref{Z242} and \eqref{V341}, respectively.
Solutions for the remaining two multiplets ${\bf (4, 4, 0)}$ and ${\bf (0, 4, 4)}$ are presented in Appendix \ref{AppB}.

One can check that all superfields listed above indeed satisfy the common constraint \eqref{mirror}. For ${\bf (1, 4, 3)}$ it is obvious.
For the rest of supermultiplets eq. \eqref{mirror} is recovered as a result of action of the appropriate covariant derivative on the basic constraints. As an instructive
example we perform this exercise for the multiplet ${\bf (3, 4, 1)}$:
\bea
    D^{(j}_{\alpha}\left[D^{i)(\alpha}V^{\beta\gamma)}\right]=0\quad \Rightarrow \quad
    D^{(j}_{\alpha}D^{i)\alpha}V^{\beta\gamma}=0.
\eea

It is also worth to point out that the chirality constraint \eqref{chiral} is also valid for some of the superfields describing
the mirror multiplets ${\bf (3, 4, 1)}$, ${\bf (4, 4, 0)}$ and ${\bf (0, 4, 4)}$, as a part of the full sets of their constraints,
\bea
    D^{i2}V^{22}=0,\qquad D^{i2}Y^{2A}=0,\qquad D^{i2}\Psi^{2A}=0.
\eea
The remaining constraints relate these chiral superfields to other (non-chiral) superfields forming a given ${\cal N}=4, d=1$ supermultiplet.
This property  allows one to construct the interacting Lagrangians as the proper superpotentials. In Section \ref{242d+341sd} such an interaction is given for
the coupled mirror ${\bf (2, 4, 2)}$ and ${\bf (3, 4, 1)}$ supermultiplets.

\subsection{Wess-Zumino action}
The Wess-Zumino (WZ) type actions (analytic superpotentials) were defined in \cite{HSS} for the ``ordinary'' multiplets as integrals over the analytic superspace,
\bea
    S^\prime_{\rm WZ}=\int  d\zeta_{\rm (A)}^{--}\,{\cal L}^{++},\qquad D^{+\alpha}{\cal L}^{++}=0. \label{WZprime}
\eea
Here ${\cal L}^{++}$ is an analytic function of harmonic analytic superfields and harmonic variables. Such a construction
is admissible for the ordinary multiplets ${\bf (0, 4, 4)}$, ${\bf (3, 4, 1)}$ and ${\bf (4, 4, 0)}$ which are described by
analytic superfields additionally  constrained by the proper harmonic conditions involving the analyticity-preserving harmonic derivative $D^{++}$.

WZ actions for the mirror superfields have the same formulation in the relevant mirror (analytic) harmonic superspace forming a subspace in the bi-harmonic superspace.
However, in this paper we prefer to construct WZ actions for mirror multiplets in the standard (ordinary) harmonic superspace.
One of the merits of this construction is that it allows a deformation to ${\rm SU}(2|1)$ supersymmetry \cite{DHSS}.

So we are going to consider an alternative construction of WZ action for mirror multiplets in the (ordinary) analytic harmonic superspace $\{\zeta_{\rm (A)}\}$\,.
Since mirror superfields carry no external harmonic charges, the only way to compensate the negative harmonic charge $-\,2$ of the invariant measure $d\zeta_{\rm (A)}^{--}$
is to include the charged objects, viz. the covariant derivatives and/or superspace coordinates.
We will try the simplest option
\bea
     S_{\rm WZ}= \int  d\zeta_{\rm (A)}^{--}\,\theta^{+}_{\alpha}D^{+}_{\beta}L^{\alpha\beta},\label{WZ}
\eea
where $L^{\alpha\beta}$ is a triplet function ($L^{\alpha\beta}=L^{\beta\alpha}$) of mirror superfields that satisfies
\bea
    D^{0}L^{\alpha\beta}=0,\qquad
    D^{++}L^{\alpha\beta}=0,\qquad D^{+}_{\gamma}D^{+\gamma}L^{\alpha\beta}=0.\label{DPDP0}
\eea
The last quadratic constraint secures the analyticity of the Lagrangian density, $D^{+\gamma} (D^{+}_{\beta}L^{\alpha\beta}) = 0\,$,
and hence the invariance of WZ action \eqref{WZ}:
\bea
    \delta S_{\rm WZ} = \int  d\zeta_{\rm (A)}^{--}\,\epsilon^{+}_{\alpha}D^{+}_{\beta}L^{\alpha\beta}
    =\int  d\zeta_{\rm (A)}^{--}\,D^{++}\left(\epsilon^{-}_{\alpha}D^{+}_{\beta}L^{\alpha\beta}\right)=0\,,
\eea
where we represented $\epsilon^{+}_{\alpha} = D^{++}\epsilon^{-}_{\alpha}$ and integrated by parts with respect to $D^{++}$.

Note that we could start from the superfield function $L^{\alpha\beta}$ having a singlet part $L$ ($L=\varepsilon_{\alpha\beta}L^{\alpha\beta}$) and
still satisfying the same constraints \eqref{DPDP0}. This part can be discarded because the relevant action is vanishing:
\bea
    \int  d\zeta_{\rm (A)}^{--}\,\theta^{+}_{\alpha}D^{+\alpha}L=\int  d\zeta_{\rm (A)}^{--}\,D^{++}\left(\theta^{-}_{\alpha}D^{+\alpha}L-L\right)=0,\qquad D^{+\gamma}\left(\theta^{-}_{\alpha}D^{+\alpha}L-L\right)=0.
\eea

\section{Spin mirror multiplet (3, 4, 1)}
In this section we treat the mirror multiplet ${\bf (3, 4, 1)}$ as semi-dynamical and construct its general WZ action.

The constraints \eqref{341} are solved by
\bea
    V^{\alpha\beta}&=&v^{\alpha\beta}+\theta^{-(\alpha}\chi^{i\beta)}u^{+}_i-\theta^{+(\alpha}\chi^{i\beta)}u^{-}_i-2i\,
    \theta^{-(\alpha}\theta^{+}_{\gamma}\dot{v}^{\beta)\gamma}+\theta^{-(\alpha}\theta^{+\beta)}C\nn
    &&-\,i\,\theta^{+\gamma}\theta^{+}_{\gamma}\theta^{-(\alpha}\dot{\chi}^{i\beta)}u^{-}_i,\label{V341}
\eea
where
\bea
    \overline{\left(v^{\alpha\beta}\right)}=-\,v_{\alpha\beta},\qquad \overline{\left(\chi^{k\alpha}\right)}=-\,\chi_{k\alpha}\,,\qquad
    \overline{\left(C\right)}=C.
\eea
The component fields transform under ${\cal N}=4, d=1$ supersymmetry  as
\bea
    \delta v^{\alpha\beta} = \epsilon^{i(\alpha} \chi^{\beta)}_{i},\quad
    \delta \chi^{i\alpha} = 2i\epsilon^{i}_{\beta}\dot{v}^{\alpha\beta} - \epsilon^{i\alpha}C,\quad
    \delta C = -\,i\epsilon_{i\alpha}\dot{\chi}^{i\alpha}, \quad \overline{\left(\epsilon^{i\alpha}\right)}=-\,\epsilon_{i\alpha}\,. \label{tr341}
\eea

Let us first construct a Fayet-Iliopoulos term as the simplest example of WZ action with $L^{\alpha\beta} \sim V^{\alpha\beta}$:
\bea
    S_{\rm FI}= \frac{b}{3}\int d\zeta_{\rm (A)}^{--}\,\theta^{+}_{\alpha}D^{+}_{\beta}\,V^{\alpha\beta} = \int dt\,{\cal L}_{\rm FI} ,\qquad {\cal L}_{\rm FI}=b\,C.\label{FI341}
\eea

A less trivial WZ action for $V^{\alpha\beta}$ is constructed according to the prescription \eqref{WZ}  as
\bea
    S_{\rm WZ} = \int dt\,{\cal L}_{\rm WZ}= \int  d\zeta_{\rm (A)}^{--}\,\theta^{+}_{\alpha}D^{+}_{\beta}\,L^{\alpha\beta}\left(V\right).\label{WZ-SF}
\eea
The zero-order component of the $\theta$-expansion of the last constraint in \eqref{DPDP0} imposes 3-dimensional Laplace equation on
the Lagrangian density $L^{\alpha\beta}\left(v\right)$:
\bea
\Delta_{(3)} L^{\alpha\beta}\left(v\right)=0,\qquad \Delta_{(3)} = \partial^{\gamma\delta}\partial_{\gamma\delta},\qquad\partial^{\gamma\delta}=\partial/\partial v_{\gamma\delta}\,.
\label{ConstrLagr}
\eea
In components, using the solution \eqref{V341} for $V^{(\alpha\beta)}$,  we obtain
\bea
    {\cal L}_{\rm WZ}=C\,{\cal U}+i\dot{v}^{\alpha\beta}{\cal A}_{\alpha\beta}+\frac{1}{2}\,{\cal R}^{\alpha\beta}\chi^{i}_{\alpha}\chi_{i\beta}\,,\label{WZL}
\eea
with
\bea
    {\cal U}\left(v\right)=\partial^{\alpha\beta}L_{\alpha\beta}\left(v\right),\;
    {\cal A}_{\alpha\beta}\left(v\right)=\varepsilon_{\alpha\gamma}\,\partial^{\gamma\delta}L_{\beta\delta}\left(v\right)
    +\varepsilon_{\beta\gamma}\,\partial^{\gamma\delta}L_{\alpha\delta}\left(v\right),\;
    {\cal R}^{\alpha\beta}\left(v\right)=\partial^{\alpha\gamma}\partial^{\beta\delta}L_{\gamma\delta}\left(v\right).\label{URAL}
\eea

Taking into account the constraint \eqref{ConstrLagr}, one finds that the quantities defined in \eqref{URAL} satisfy the conditions
\bea
    \partial_{\alpha\beta}\,{\cal U}={\cal R}_{\alpha\beta}\,,\quad \Delta_{(3)}\,{\cal U}=\Delta_{(3)} {\cal R}_{\alpha\beta} = 0,\quad
    \partial_{\alpha\beta}\,{\cal A}_{\gamma\delta}-\partial_{\gamma\delta}\,{\cal A}_{\alpha\beta}
    =\varepsilon_{\alpha\gamma}\,{\cal R}_{\beta\delta}+\varepsilon_{\beta\delta}\,{\cal R}_{\alpha\gamma}\,. \label{ConstrOnWZ}
\eea
Swapping, in the Lagrangian \eqref{WZL} and the constraints \eqref{ConstrOnWZ}, the indices  $\alpha, \beta$ and $i,j\,$,  we obtain just WZ Lagrangian
for the ordinary multiplet ${\bf (3, 4, 1)}$ constructed in \cite{HSS}. Thus the above formulas yield the correct form of WZ action
for the mirror multiplet ${\bf (3, 4, 1)}$.

Eliminating the fermionic fields in \eqref{WZL} by their equations of motion
we pass to the Hamiltonian system with:
\bea
    H=\lambda^{\alpha\beta}\pi_{\alpha\beta}-C\,{\cal U},
\eea
where $\lambda^{\alpha\beta}$ and $C$ are treated as Lagrange multipliers.
The second class Hamiltonian constraints of the system are then given by
\bea
    \pi_{\alpha\beta}=p_{\alpha\beta}-i\,{\cal A}_{\alpha\beta}\approx 0,\qquad  {\cal U}\approx 0.\label{cons}
\eea
Note that the last constraint is a secondary one for the primary constraint $p_C \approx 0$.

The matrix formed by Poisson brackets of the constraints \eqref{cons} is not degenerate:
\bea
{\rm det}
\begin{pmatrix}
\left\lbrace \pi_{\alpha\beta},\pi_{\gamma\delta} \right\rbrace_{\rm PB}&  \left\lbrace \pi_{\alpha\beta},{\cal U} \right\rbrace_{\rm PB} \\
 \left\lbrace {\cal U},\pi_{\gamma\delta} \right\rbrace_{\rm PB} &  0\\
\end{pmatrix}
\neq 0,
\eea
and hence  we can pass to Dirac brackets. Calculating the inverse matrix of these constraints, we find the Dirac brackets in the form
\bea
    \left\lbrace v_{\alpha\beta}, v_{\gamma \delta}\right\rbrace=\frac{i
    \left(\varepsilon_{\alpha\gamma}{\cal R}_{\beta\delta}+\varepsilon_{\beta\delta}{\cal R}_{\alpha\gamma}\right)}{2\,{\cal R}^{\lambda\mu}{\cal R}_{\lambda\mu}}\,.
\eea
One can check that
\bea
    \left\lbrace v_{\alpha\beta}, {\cal U}\right\rbrace = \frac{i
    \left({\cal R}_{\beta\gamma}{\cal R}^{\gamma}_{\alpha}+{\cal R}_{\alpha\gamma}{\cal R}^{\gamma}_{\beta}\right)}{2\,{\cal R}^{\lambda\mu}{\cal R}_{\lambda\mu}}=0.
\eea
The constraint ${\cal U}\approx 0$ kills one degree of freedom in the triplet $v^{\alpha\beta}$,
so this triplet effectively describes a $2$-dimensional surface embedded in $\mathbb{R}^3$. 

\subsection{Non-commutative plane}
Let us consider the simplest solution of the Laplace equation $\Delta_{(3)}\,{\cal U}=0$,
\bea
    {\cal U}=\frac{c-y}{2}\,,\qquad c={\rm const},\label{nonc}
\eea
where
\bea
    v_{12}=y,\qquad v_{11}=-\,\sqrt{2}\,u,\qquad v_{22}=\sqrt{2}\,\bar{u}.
\eea
It corresponds to the following choice of the triplet function $L^{\alpha\beta}$
\bea
    L^{11}=0,\qquad L^{22}=0,\qquad L^{12}=\frac{1}{4}\left(y^2-u\bar{u}-2cy\right).
\eea

The relevant Lagrangian is then written as
\bea
    {\cal L}_{\rm WZ}=\frac{i}{2}\left(u\dot{\bar{u}}-\dot{u}\bar{u}\right)+\frac{C}{2}\left(c-y\right)-\frac{1}{4}\,\chi^{i}_{1}\chi_{i2}\,.\label{NCLagr}
\eea
It is straightforward to check that it is invariant off shell under the following ${\cal N}=4$ supersymmetry transformations \eqref{tr341}.
The Lagrangian \eqref{NCLagr} is none other than ${\cal N}=4$, $d=1$ supersymmetrization of the $d=1$ WZ Lagrangian describing the lowest level of the planar
Landau model (see, e.g., \cite{Plan} for a review). In particular, besides the standard phase ${\rm U}(1)_R$ transformations, it is invariant under the so called ``magnetic translations''
\be
\delta u = \lambda\,, \qquad \delta \bar u = \bar\lambda\,, \label{MagTr}
\ee
with $\lambda$ being a complex parameter. It is worth noting that the analogous system for the ``ordinary'' multiplet ${\bf(3, 4, 1)}$ is described by the action \eqref{WZprime}
with ${\cal L}^{++} \sim V^{++} + c^{--}\left(V^{++}\right)^2$, where the analytic superfield $V^{++}$ satisfies the constraint $D^{++}V^{++} = 0$ and $c^{--} = c^{ik}u^-_iu^-_k$.
Fixing ${\rm SU}(2)_{\rm L}$ frame as $c^{11} = c^{22} =0, \;c^{12} \neq 0,$ and making the appropriate redefinitions, one arrives at the WZ Lagrangian \eqref{NCLagr} with the
swapped ${\rm SU}(2)_{{\rm L}, {\rm R}}$ indices \footnote{It is curious that the magnetic translations \eqref{MagTr} are realized in the ``ordinary'' description
as $\delta V^{++} = \lambda^{ik}u^+_iu^+_k\,, c^{ik}\lambda_{ik} = 0$. There is the corresponding superfield realization of these transformations in the mirror description too.}.

The matrix of the second-class constraints in this case takes the very simple non-degenerate form
\bea
\begin{pmatrix}
0 & i & 0 & 0\\
\;-\,i & 0 & 0 & 0\\
0 & 0 & 0 & 1/2\;\\
0 & 0 & -\,1/2 & 0
\end{pmatrix}.
\eea
The Dirac brackets are
\bea
    \left\lbrace u,\bar{u}\right\rbrace = i,\qquad \left\lbrace y, {u}\right\rbrace = 0,\qquad \left\lbrace y,\bar{u}\right\rbrace = 0\,.\label{noncbr}
\eea
The complex field $u$ describes a non-commutative plane in $\mathbb{R}^3$, while the third coordinate (component) $y$, perpendicular to this plane,
takes the constant value $y=c$.

In \cite{FIL2012}, the fuzzy sphere solution was considered (for the ``ordinary'' ${\bf (3, 4, 1)}$ multiplet) as a solution of the 3-dimensional Laplace equation:
\bea
    {\cal U} \sim \frac{1}{\sqrt{y^2+2u\bar{u}}}\,,\qquad \left(\partial^2_y +2\,\partial_{u}\partial_{\bar u}\right){\cal U}=0.
\eea
The non-commutative plane was not considered, so here we fill this gap. The non-commutative plane is the planar limit of the fuzzy sphere.
We choose the suitable solution by shifting the center of the sphere as
\bea
    {\cal U}=\frac{1}{2}\left[c+R-\frac{R^2}{\sqrt{\left(y-R\right)^2+2u\bar{u}}}\right],
\eea
with $R$ being the radius. In the limit $R\rightarrow\infty$ we recover the plane solution \eqref{nonc}.

Note that an actual effect of considering the
${\cal N}=4, d=1$ WZ Lagrangians in the present context is manifested  while coupling them to the matter ${\bf (1, 4, 3)}$ multiplet, where these Lagrangians give rise to additional
on-shell potential terms and Yukawa-type couplings (see the next Section).

\section{The mirror system (3, 4, 1) -- (1, 4, 3) and Nahm equations}
As an instructive example we consider the simplest coupling of the semi-dynamical mirror multiplet ${\bf (3, 4, 1)}$
and the mirror multiplet ${\bf (1, 4, 3)}$. In fact, we consider the same model as the one constructed in \cite{FIL2012}, but in terms of mirror superfields.
Swapping $\alpha, \beta$ and $i,j$ indices, we reproduce its Lagrangian and Nahm equations associated with its Hamiltonian formulation. In the end we will consider
a deformation to ${\rm SU}(2|1)$ supersymmetry.

\subsection{Dynamical mirror multiplet (1, 4, 3)}
The duality between two ${\bf (1, 4, 3)}$ multiplets was studied in \cite{IKP} and, later on, in \cite{DelIva2007}. It was shown there that, inserting
the constraints \eqref{143} into the invariant action with a superfield Lagrangian multiplier and integrating the superfield $X$ out,
we obtain the action and constraint for the ``ordinary'' multiplet ${\bf (1, 4, 3)}$ described by the former superfield Lagrangian multiplier.
In our terminology, the mirror ${\bf (1, 4, 3)}$ multiplet is described just by the superfield $X$ with the constraints \eqref{143}.

Solving this constraint, we obtain
\bea
    X=x-\theta^{-}_{\alpha}\psi^{i\alpha}u^{+}_i+\theta^{+}_{\alpha}\psi^{i\alpha}u^{-}_i+\theta^{-}_{(\alpha}\theta^{+}_{\beta)}A^{\alpha\beta}
    +i\,\theta^{-}_{\alpha}\theta^{+\alpha}\dot{x}+i\,\theta^{+\alpha}\theta^{+}_{\alpha}\theta^{-}_{\beta}\dot{\psi}^{i\beta}u^{-}_i,\label{X143}
\eea
where
\bea
    \overline{\left(x\right)}=x,\qquad  \overline{\left(\psi^{i\alpha}\right)}=\psi_{i\alpha}\,,\qquad \overline{\left(A^{\alpha\beta}\right)}=-\,A_{\alpha\beta}\,.
\eea
Supersymmetry transformations are
\bea
    \delta x = \epsilon_{i\alpha}\psi^{i\alpha},\qquad
    \delta \psi^{i\alpha} = \epsilon^i_{\beta}A^{\alpha\beta}+i\epsilon^{i\alpha}\dot{x},\qquad
    \delta A^{\alpha\beta} = 2i\epsilon^{i(\alpha}\dot{\psi}^{\beta)}_i.\label{tr143}
\eea
The kinetic Lagrangian for the mirror multiplet ${\bf (1, 4, 3)}$ is given by the superfield action \cite{DelIva2007}
\bea
    S_{\rm kin.}=\frac{1}{2}\int d\zeta_{\rm H}\,f\left(X\right) = \int dt\,{\cal L}_{\rm kin}.
\eea
The component Lagrangian reads
\bea
    {\cal L}_{\rm kin.}=g\left[\frac{\dot{x}^2}{2}+\frac{i}{2}\,\psi^{i\alpha}\dot{\psi}_{i\alpha}
    -\frac{A^{\alpha\beta}A_{\alpha\beta}}{4}\right]-\frac{1}{4}\,g^{\prime}A^{\alpha\beta}\psi^i_{\alpha}\psi_{i\beta}
    -\frac{1}{24}\,g^{\prime\prime}\psi^{i}_{\alpha}\psi_{i\beta}\psi^{j\alpha}\psi^{\beta}_{j}\,, \label{kin}
\eea
where $g:=g\left(x\right)=f^{\prime\prime}\left(x\right)$.

The relevant Fayet-Iliopoulos term is defined as
\bea
    S_{\rm FI}= b^{\alpha\beta}\int d\zeta_{\rm (A)}^{--}\,\theta^{+}_{\alpha}D^{+}_{\beta}\,X = \int dt{\cal L}_{\rm FI},\qquad {\cal L}_{\rm FI}=b^{\alpha\beta}A_{\alpha\beta}\,.\label{FI143}
\eea
\subsection{Couplings and total Lagrangian}
The total Lagrangian is a sum of three Lagrangians:
\bea
    {\cal L}_{\rm tot.}={\cal L}_{\rm kin.}+{\cal L}_{\rm WZ}+{\cal L}_{\rm int.}\,.
\eea
The kinetic and WZ Lagrangians are given by \eqref{kin} and \eqref{WZL}. We could add the Fayet-Iliopoulos Lagrangians \eqref{FI341} and \eqref{FI143},
but they bring only potential terms and therefore have no impact on the structure of brackets, which is our main subject here. The Lagrangian ${\cal L}_{\rm int.}$ describes an interaction
of the mirror ${\bf (1, 4, 3)}$ and ${\bf (3, 4, 1)}$ multiplets,
\bea
    S_{\rm int.} = \int dt\,{\cal L}_{\rm int.} = \frac{\mu}{2}\int d\zeta^{--}_{\rm A}\,h^{++}, \label{int-SF}
\eea
where $h^{++}$ is analytic. From ref. \cite{FIL2012} we know that the interaction term in the ``ordinary'' case involves both
dynamical and semi-dynamical superfields linearly. Obviously, the same should be true for their mirror counterparts $X$ and $V^{\alpha\beta}$. Supposing this,  we find that
the correct ansatz for $h^{++}$ is
\bea
    &&h^{++}=\theta^{+\alpha}V_{\alpha\beta}\left(D^{+\beta}X\right)+\frac{1}{3}\,
    \theta^{+\alpha}X\left(D^{+\beta}V_{\alpha\beta}\right)+\frac{1}{3}\,\theta^{-}_{\gamma}\theta^{+\gamma}\left(D^{+\alpha}V_{\alpha\beta}\right)\left(D^{+\beta}X\right),\nn
    &&D^{+\gamma}h^{++}=0,\qquad D^{++}h^{++}\neq 0.\label{coupling143}
\eea
Now one can directly check that the action is invariant:
\bea
    &&\delta S_{\rm int.}= \frac{\mu}{2}\int d\zeta^{--}_{\rm A}\,\delta{h}^{++}= \frac{\mu}{2}\int d\zeta^{--}_{\rm A}\,D^{++}\delta h=0,\qquad D^{+\gamma}\delta h=0,\nn
    &&\delta h=\left[\epsilon^{-\alpha}V_{\alpha\beta}\left(D^{+\beta}X\right)+\frac{1}{3}\,
    \epsilon^{-\alpha}X\left(D^{+\beta}V_{\alpha\beta}\right)+\frac{1}{3}\,\epsilon^{-}_{\gamma}\theta^{-\gamma}\left(D^{+\alpha}V_{\alpha\beta}\right)\left(D^{+\beta}X\right)\right].
\eea
The  component Lagrangian is found to be
\bea
    {\cal L}_{\rm int.}=\frac{\mu}{2}\left(x\,C+A^{\alpha\beta}v_{\alpha\beta}-\psi^{i\alpha}\chi_{i\alpha}\right).\label{intL}
\eea
Eliminating the auxiliary fields $\chi^{i\alpha}$ and $A^{\alpha\beta}$ by their equations of motion, we obtain the total Lagrangian:
\bea
    {\cal L}_{\rm tot.}&=&g\left[\frac{\dot{x}^2}{2}+\frac{i}{2}\,\psi^{i\alpha}\dot{\psi}_{i\alpha}\right]
    +\frac{\mu^2\,v^{\alpha\beta}v_{\alpha\beta}}{4g}-\frac{\mu\,g^{\prime}v_{\alpha\beta}}{4g}\,\psi^{i\alpha}\psi_{i}^{\beta}
    +i\dot{v}^{\alpha\beta}{\cal A}_{\alpha\beta}-\frac{\mu^2{\cal R}^{\alpha\beta}\psi^{i}_{\alpha}\psi_{i\beta}}{4{\cal R}^{\gamma\delta}{\cal R}_{\gamma\delta}}\nn
    &&-\,\frac{1}{24}\left[g^{\prime\prime}-\frac{3\left(g^{\prime}\right)^2}{2g}\right]\psi^{i}_{\alpha}\psi_{i\beta}\psi^{j\alpha}\psi^{\beta}_{j}
    +C\left(\frac{\mu\,x}{2}+{\cal U}\right).
    \label{total}
\eea

\subsection{Nahm equations}
The Hamiltonian corresponding to \eqref{total} reads
\bea
    H&=&\frac{p^2}{2g}-\frac{\mu^2\,v^{\alpha\beta}v_{\alpha\beta}}{4g}+\frac{\mu\,g^{\prime}v_{\alpha\beta}}{4g}\,\psi^{i\alpha}\psi_{i}^{\beta}
    +\frac{\mu^2{\cal R}^{\alpha\beta}\psi^{i}_{\alpha}\psi_{i\beta}}{4{\cal R}^{\gamma\delta}{\cal R}_{\gamma\delta}}
    +\frac{1}{24}\left[g^{\prime\prime}-\frac{3\left(g^{\prime}\right)^2}{2g}\right]\psi^{i}_{\alpha}\psi_{i\beta}\psi^{j\alpha}\psi^{\beta}_{j}\nn
    &&+\,\lambda^{\alpha\beta}\pi_{\alpha\beta}-C\left(\frac{\mu\,x}{2}+{\cal U}\right)+\tilde{\lambda}^{i\alpha}\tilde{\pi}_{i\alpha}\,.
\eea
The relevant Hamiltonian constraints  are
\bea
    \pi_{\alpha\beta}=p_{\alpha\beta}-i\,{\cal A}_{\alpha\beta}\approx 0,\qquad h=\frac{\mu\,x}{2}
    +{\cal U} \approx 0,\qquad \tilde{\pi}_{i\alpha}=p_{i\alpha}+\frac{i}{2}\,g\,\psi_{i\alpha}\,.
\eea
We observe that the ${\cal N}=4$ supersymmetric coupling to the mirror dynamical multiplet modifies the previous constraint ${\cal U} \approx 0$ as
\bea
    h={\cal U} + \frac{\mu\,x}{2}\approx 0.\label{h}
\eea
It relates one degree of freedom of the spin variables $v^{\alpha\beta}$ to the dynamical bosonic field $x$.

The Dirac brackets are calculated as:
\bea
    &&\left\lbrace x, p\right\rbrace = 1,\qquad
    \left\lbrace v_{\alpha\beta}, v_{\gamma \delta}\right\rbrace=\frac{i\left(\varepsilon_{\alpha\gamma}{\cal R}_{\beta\delta}
    +\varepsilon_{\beta\delta}{\cal R}_{\alpha\gamma}\right)}{2\,{\cal R}^{\lambda\mu}{\cal R}_{\lambda\mu}}\,,\qquad
    \left\lbrace p, v_{\alpha\beta}\right\rbrace = \frac{\mu\,{\cal R}_{\alpha\beta}}{2\,{\cal R}^{\lambda\mu}{\cal R}_{\lambda\mu}}\,,\nn
    &&\left\lbrace \psi^{i\alpha}, \psi_{j\beta}\right\rbrace = -\,\frac{i}{g}\,\delta^i_j\delta^{\alpha}_{\beta}\,,\qquad
    \left\lbrace p, \psi^{i\alpha}\right\rbrace = \frac{1}{2g}\,g^{\prime}\psi^{i\alpha}.
    \label{brackets}
\eea
In a complete analogy with the results of ref. \cite{FIL2012} for the ``ordinary'' multiplets,  the triplet of spin variables $v^{\alpha\beta}$ describes
2-dimensional surface in $\mathbb{R}^3$ defined by the equations:
\bea
    \left\lbrace v_{\alpha\beta},v_{\gamma \delta}\right\rbrace = \frac{i}{\mu}\left(\varepsilon_{\alpha\gamma}
    \left\lbrace p, v_{\beta\delta} \right\rbrace +\varepsilon_{\beta\delta}\left\lbrace p, v_{\alpha\gamma}\right\rbrace\right).\label{NahmEqs}
\eea
These are just famous Nahm equations \cite{Nahm} \footnote{To be more exact, it is some generalization of them (see, e.g., \cite{GenNahm} and references therein).} and they can be put in the standard form as
\bea
    \left\lbrace p, v_{c} \right\rbrace = \frac{1}{2}\,\varepsilon_{abc}\left\lbrace v_{a},v_{b}\right\rbrace ,\qquad v_{\alpha\gamma} \rightarrow  \frac{v_a}{\mu}\,,\qquad a=1,2,3.
\eea
Here $x$ plays the role of evolution parameter and $p$ appears as a derivation with respect to the latter. Thus we obtained a model equivalent to the model constructed
earlier in \cite{FIL2012}. To establish the exact equivalence, we need to interchange the SU$(2)$
indices as $i, j\leftrightarrow \alpha, \beta$.

Let us consider as an example Nahm equations for the non-commutative plane \eqref{nonc}. The constraint \eqref{h} implies that
\bea
    y=\mu\,x+c.
\eea
We obtain the same Dirac brackets \eqref{noncbr} for the spin variables.
The relevant Nahm equations are written as
\bea
    \left\lbrace u,\bar{u}\right\rbrace = \frac{i}{\mu}\left\lbrace y,p\right\rbrace =\frac{i}{\mu}\,\partial_x  y,\qquad
    \left\lbrace y,\bar{u}\right\rbrace = -\,\frac{i}{\mu}\left\lbrace \bar{u},p\right\rbrace =0,\qquad \left\lbrace y, u\right\rbrace =\frac{i}{\mu}\left\lbrace u,p\right\rbrace = 0,\label{nplane}
\eea
where the perpendicular coordinate $y$ is directly related to the dynamical component $x$.

The resume of this subsection is that the ${\cal N}=4, d=1$ supersymmetric coupling of the mirror dynamical ${\bf (1, 4, 3)}$ and semi-dynamical
${\bf (3, 4, 1)}$ multiplets reveals no new features compared to its analog for the ``ordinary'' multiplets of this kind. All the results, suggestions and conjectures of \cite{FIL2012}
apply for the mirror multiplets as well. In particular, just the Nahm equations of the type discussed above ensure the correct closure  of  ${\cal N}=4$ supercharges and the Hamiltonian
in both the classical and the quantum cases.

\subsection{${\rm SU}(2|1)$ supersymmetry}
We limit our consideration of the deformed ${\rm SU}(2|1)$, $d=1$ supersymmetry  by the component level, following refs. \cite{DSQM,SKO,DHSS}.
As was shown in \cite{DHSS}, deformed multiplets and their mirror counterparts cease to be equivalent after such a deformation. In particular, WZ Lagrangians for the
multiplet ${\bf (4, 4, 0)}$ can be constructed only if it belongs to the mirror type.
A similar situation is expected for the multiplets ${\bf (3, 4, 1)}$. Until now we were able to construct self-consistent  ${\rm SU}(2|1)$ invariant
WZ Lagrangians only for the mirror multiplets.

For a start, the centrally-extended superalgebra $su(2|1)\oplus u(1)$ is defined by the following non-vanishing (anti)commutator
\footnote{The deformed supercharges originally defined in \cite{DSQM} correspond to $Q^{i}:=Q^{i1}$, $\bar{Q}_{j}:=-\,Q_{j1}$\,.}:
\bea
    &&\left\lbrace Q^{i}_{\beta}, Q_{j}^{\alpha}\right\rbrace =  2\delta^i_j\delta^\alpha_\beta \left(H-mF\right)-2m
    \left(\sigma_3\right)^\alpha_\beta I^i_j ,\qquad\left[I^i_j,  I^k_l\right]
    = \delta^k_j I^i_l - \delta^i_l I^k_j\,,\nn
    &&\left[I^i_j, Q^{k\alpha}\right]
    = \delta^k_j Q^{i\alpha} - \frac{1}{2}\,\delta^i_j Q^{k\alpha},\qquad
    \left[F, Q^{i\alpha}\right]=\frac{1}{2}\left(\sigma_3\right)^\alpha_\beta Q^{i\beta},\nn
    &&\left[H, Q^{i}_{\beta}\right] = 0,\qquad \left[H, F\right] = 0,\qquad \left[H, I^i_j\right] = 0,\qquad \left[I^i_j, F\right] = 0.\label{su21}
\eea
Here $\sigma_3$ is the standard Pauli matrix:
\bea
    \left(\sigma_3\right)^{1}_{1}=-\left(\sigma_3\right)^{2}_{2}=1
\eea
and the Hamiltonian $H$ is treated as a central charge operator commuting with all other generators. The superalgebra \eqref{su21}
contains additional bosonic generators $I^i_j$ and $F$ which form the subalgebra $su(2)_{\rm L} \oplus u(1)_{\rm R}$.
Hence the equivalence between ordinary and mirror multiplets cannot be valid for SU$(2|1)$ supersymmetry since swapping of the ${\rm SU}(2)_{\rm L}$
and ${\rm SU}(2)_{\rm R}$ indices yields a different superalgebra.

We skip details of solving superfield constraints, and proceed to the component transformations and Lagrangians.

For the dynamical mirror multiplet ${\bf (1, 4, 3)}$ the deformation of the transformation laws \eqref{tr143} amounts to
\bea
    &&\delta x = \epsilon_{i\alpha}\psi^{i\alpha},\qquad
    \delta A^{\alpha\beta} = 2\epsilon^{i(\alpha}\left[i\dot{\psi}_{i}^{\beta)}
    +m\left(\sigma_3\right)_{\gamma}^{\beta)}\psi_{i}^{\gamma}\right]+m\left(\sigma_3\right)^{\alpha\beta}\epsilon_{i\gamma}\psi^{i\gamma},\nn
    &&\delta \psi^{i\alpha} = \epsilon^i_{\beta}A^{\alpha\beta}+i\epsilon^{i\alpha}\dot{x}.\label{tr143d}
\eea
The deformed kinetic Lagrangian invariant under these transformations is as follows
\bea
    {\cal L}_{\rm kin.}&=&g\left[\frac{\dot{x}^2}{2}+\frac{i}{2}\,\psi^{i\alpha}\dot{\psi}_{i\alpha}-\frac{A^{\alpha\beta}A_{\alpha\beta}}{4}
    -\frac{m}{2}\left(\sigma_3\right)^{\alpha}_{\beta}\psi^{i\beta}\psi_{i\alpha}\right]
    -\frac{1}{4}\,g^{\prime}A^{\alpha\beta}\psi^i_{\alpha}\psi_{i\beta}+\frac{m}{2}\,f^{\prime}\left(\sigma_3\right)^{\alpha\beta}\,A_{\alpha\beta}\nn
    &&-\,\frac{1}{24}\,g^{\prime\prime}\psi^{i}_{\alpha}\psi_{i\beta}\psi^{j\alpha}\psi^{\beta}_{j}\,.
\eea
The transformations \eqref{tr341} of the spin multiplet ${\bf (3, 4, 1)}$ are deformed as
\bea
    &&\delta v^{\alpha\beta} = \epsilon^{i(\alpha} \chi^{\beta)}_{i},\quad
    \delta C = -\,i\epsilon_{i\alpha}\dot{\chi}^{i\alpha},\quad
    \delta \chi^{i\alpha} = 2\epsilon^{i}_{\beta}\left[i\dot{v}^{\alpha\beta}+m\left(\sigma_3\right)_{\gamma}^{(\alpha} v^{\beta)\gamma}\right]
    - \epsilon^{i\alpha}C.\label{tr341d}
\eea
The deformed WZ Lagrangian is then given by
\bea
    {\cal L}_{\rm WZ}=C\,{\cal U}+i\dot{v}^{\alpha\beta}{\cal A}_{\alpha\beta}
    +\frac{1}{2}\,{\cal R}^{\alpha\beta}\chi^{i}_{\alpha}\chi_{i\beta}+m\left(\sigma_3\right)^{\alpha}_{\beta}v^{\beta\gamma}{\cal A}_{\alpha\gamma}\,,\label{WZd}
\eea
where the quantities ${\cal U}, {\cal A}_{\alpha\beta}$ and ${\cal R}^{\alpha\beta}$ are still defined  according to eqs. \eqref{URAL} and \eqref{ConstrOnWZ}, with the ``prepotential''
$L^{\alpha\beta}$ satisfying the 3-dim Laplace equation \eqref{ConstrLagr}. The ${\rm SU}(2|1)$ invariance requires the deformed Lagrangian \eqref{WZd}
to be invariant under U$(1)_{\rm R}$ symmetry
\footnote{If we pass to the Hamiltonian $\tilde{H}:=H-mF$ \cite{SKO}, the U$(1)_{\rm R}$ generator $F$ becomes an external automorphism generator
and we can withdraw the condition \eqref{U1}. Passing to the new basis requires a redefinition of the component fields as
\bea
    &&v^{\alpha\beta} \rightarrow \frac{1}{2}\left[v^{\alpha\gamma}e^{im\left(\sigma_3\right)^{\beta}_{\gamma}}+v^{\gamma\beta}e^{im\left(\sigma_3\right)^{\alpha}_{\gamma}}\right],\qquad
    \chi^{i\alpha}\rightarrow \chi^{i\gamma}e^{\frac{i}{2}m\left(\sigma_3\right)^{\alpha}_{\gamma}},\qquad
    C\rightarrow C,\nn
    &&A^{\alpha\beta} \rightarrow \frac{1}{2}\left[A^{\alpha\gamma}e^{im\left(\sigma_3\right)^{\beta}_{\gamma}}+A^{\gamma\beta}e^{im\left(\sigma_3\right)^{\alpha}_{\gamma}}\right],\qquad
    \psi^{i\alpha}\rightarrow \psi^{i\gamma}e^{\frac{i}{2}m\left(\sigma_3\right)^{\alpha}_{\gamma}},\qquad
    x\rightarrow x.\nonumber
\eea
In the new basis the Lagrangian \eqref{WZd} gets undeformed and the Lagrangian \eqref{intL} stays undeformed, whereas the conditions \eqref{U1} can be preserved (in this case the Lagrangian
is invariant under the external $F$ automorphisms) or dismissed (in this case no extra U$(1)$ invariance is present). In the second case the inverse transformation to
the original variables would yield a generalization of the Lagrangian \eqref{WZd} by some $t$-dependent terms breaking the invariance under time-translations
(with $H$ as the corresponding generator). The requirement of absence of such terms leads, once again, to eqs. \eqref{U1}.\label{footnoteU1}}, which imposes additional conditions
on \eqref{WZd}:
\bea
    m\left(\sigma_3\right)^{\gamma}_{\lambda}v^{\delta\lambda}{\cal R}_{\gamma\delta}=0,\qquad
    m\left[\left(\sigma_3\right)^{\lambda}_{\delta}v^{\delta\gamma}\partial_{\gamma\lambda}{\cal A}_{\alpha\beta}+\left(\sigma_3\right)^{\gamma}_{\alpha}{\cal A}_{\beta\gamma}\right]=0.\label{U1}
\eea
The necessity of these conditions for invariance of the Lagrangian \eqref{WZd} under the deformed transformations \eqref{tr341d} can be directly checked.

Surprisingly, the interaction term \eqref{intL} is invariant under the deformed transformations \eqref{tr143d} and \eqref{tr341d} as it stands, {\it i.e.} it stays undeformed.

Finally, the total Lagrangian reads
\bea
    {\cal L}_{\rm tot.}&=&g\left[\frac{\dot{x}^2}{2}+\frac{i}{2}\,\psi^{i\alpha}\dot{\psi}_{i\alpha}
    -\frac{m}{2}\left(\sigma_3\right)^{\alpha}_{\beta}\psi^{i\beta}\psi_{i\alpha}\right]+\frac{1}{4g}\left[\mu\,v^{\alpha\beta}
    +m\,f^{\prime}\left(\sigma_3\right)^{\alpha\beta}\right]\left[\mu\,v_{\alpha\beta}+m\,f^{\prime}\left(\sigma_3\right)_{\alpha\beta}\right]\nn
    &&-\,\frac{g^{\prime}}{4g}\left[\mu\,v_{\alpha\beta}+m\,f^{\prime}\left(\sigma_3\right)_{\alpha\beta}\right]\psi^{i\alpha}\psi_{i}^{\beta}
    +i\dot{v}^{\alpha\beta}{\cal A}_{\alpha\beta}-\frac{\mu^2{\cal R}^{\alpha\beta}\psi^{i}_{\alpha}\psi_{i\beta}}{4{\cal R}^{\gamma\delta}{\cal R}_{\gamma\delta}}
    +m\left(\sigma_3\right)^{\alpha}_{\beta}v^{\beta\gamma}{\cal A}_{\alpha\gamma}\nn
    &&-\,\frac{1}{24}\left[g^{\prime\prime}-\frac{3\left(g^{\prime}\right)^2}{2g}\right]\psi^{i}_{\alpha}\psi_{i\beta}\psi^{j\alpha}\psi^{\beta}_{j}
    +C\left(\frac{\mu\,x}{2}+{\cal U}\right).
    \label{totald}
\eea

After passing to the Hamiltonian formalism, the brackets \eqref{brackets} and Nahm equations \eqref{NahmEqs} keep their form. This is due to the fact that new terms
$\sim m$ and $\sim m^2$ appear without time derivatives, {\it i.e.} they all are potential terms.

The ${\rm SU}(2|1)$ supercharges for the simplest free Lagrangian corresponding to $f=x^2/2$ and $g=1$ are written as
\bea
    Q^{i\alpha}= i\,p\,\psi^{i\alpha} + \left[\mu\,v^{\alpha}_{\gamma}
    +m\,x\left(\sigma_3\right)^{\alpha}_{\gamma}\right]\psi^{i\gamma}.
\eea
The bracket for the fermionic fields is simplified to
\bea
    \left\lbrace \psi^{i\alpha}, \psi_{j\beta}\right\rbrace = -\,i\,\delta^i_j\delta^{\alpha}_{\beta}\,.\label{psipsi}
\eea
Taking into account this bracket, we obtain that \footnote{Here we deal with the classical (anti)commutators generated by Dirac brackets,
 when the right-hand sides in the superalgebra \eqref{su21} are multiplied by $-i$.}
\bea
    \left\lbrace Q^{i}_{\beta}, Q_{j}^{\alpha}\right\rbrace_{\rm cl.} &=&-\,i\,\delta^i_j\delta^\alpha_\beta\left(p^2-\frac{1}{2}\left[\mu\,v^{\gamma\delta}
    +m\,x\left(\sigma_3\right)^{\gamma\delta}\right]\left[\mu\,v_{\gamma\delta}+m\,x\left(\sigma_3\right)_{\gamma\delta}\right]+\mu\left\lbrace p,v^{\gamma\delta}\right\rbrace\psi^{k}_{\gamma}\psi_{k\delta}\right)\nn
    &&-\,\frac{i}{2}\,\delta^i_j\,m\left(\sigma_3\right)^{\delta}_{\gamma}\psi^{k\gamma}\psi_{k\delta}+im\left(\sigma_3\right)^{\alpha}_{\beta}\psi^{i\gamma}\psi_{j\gamma}\nn
    &&+\,\underline{i\mu\left\lbrace p,v^{\alpha\delta}\right\rbrace\psi^{i}_{\delta}\psi_{j\beta}+i\mu\left\lbrace p,v_{\beta\gamma}\right\rbrace\psi^{i\alpha}\psi^{\gamma}_{j}-\mu^2\left\lbrace v_{\beta\gamma},v^{\alpha\delta}\right\rbrace\psi^{i\gamma}\psi_{j\delta}}\,.
\eea
The underlined expression vanishes due to the Nahm equations \eqref{NahmEqs}.
Thus, the supercharges close on the following bosonic generators:
\bea
   H-mF &=&\frac{p^2}{2}-\frac{1}{4}\left[\mu\,v^{\alpha\beta}
    +m\,x\left(\sigma_3\right)^{\alpha\beta}\right]\left[\mu\,v_{\alpha\beta}+m\,x\left(\sigma_3\right)_{\alpha\beta}\right]
    +\frac{\mu}{2}\left\lbrace p,v^{\alpha\beta}\right\rbrace\psi^{k}_{\alpha}\psi_{k\beta}\nn
    &&+\,\frac{m}{4}\left(\sigma_3\right)^{\alpha}_{\beta}\psi^{k\beta}\psi_{k\alpha}\,,\nn
    I^{i}_{j}&=&\frac{1}{2}\,\psi^{i\alpha}\psi_{j\alpha}\,.
\eea
Then the generator $\tilde{H} := H-mF$ can be divided into
\bea
    H&=&\frac{p^2}{2}-\frac{1}{4}\left[\mu\,v^{\alpha\beta}
    +m\,x\left(\sigma_3\right)^{\alpha\beta}\right]\left[\mu\,v_{\alpha\beta}+m\,x\left(\sigma_3\right)_{\alpha\beta}\right]
    +\frac{\mu}{2}\left\lbrace p,v^{\alpha\beta}\right\rbrace\psi^{k}_{\alpha}\psi_{k\beta}\nn
    &&+\,\frac{m}{2}\left(\sigma_3\right)^{\alpha}_{\beta}\psi^{k\beta}\psi_{k\alpha}-m\left(\sigma_3\right)^{\alpha}_{\beta}v^{\beta\gamma}{\cal A}_{\alpha\gamma}\,,\nn
    F&=&\left(\sigma_3\right)^{\alpha}_{\beta}\left[\frac{1}{4}\,\psi^{k\beta}\psi_{k\alpha}-v^{\beta\gamma}{\cal A}_{\alpha\gamma}\right].
\eea
The term $\sim\left(\sigma_3\right)^{\alpha}_{\beta}v^{\beta\gamma}{\cal A}_{\alpha\gamma}$ enters as a part of both $H$ and $F$, but it is absent in their combination
$\tilde{H} = H-mF$. Due to the presence of this term the correct commutators of $H$ and $F$ with supercharges are not guaranteed
by the Nahm equations and the bracket \eqref{psipsi} only. One also needs to make use of the whole set of the Dirac brackets \eqref{brackets}
and to keep in mind the conditions \eqref{U1}. Thus the Nahm equations \eqref{NahmEqs} and the fermionic bracket \eqref{psipsi} alone
suffice to provide relations for the $su(2|1)$ superalgebra without central charge \cite{SKO}, in which the bosonic generator $\tilde{H}=H-mF$ plays the role of
the Hamiltonian, while the generator $F$ corresponds to the external automorphisms under which the Lagrangian is not obliged to be invariant
(see discussion in the footnote \ref{footnoteU1}).

\section{Coupling with a chiral multiplet}\label{242d+341sd}
Here we construct the superfield and component couplings of the mirror multiplets ${\bf (2,4,2)}$ and ${\bf (3,4,1)}$. The first multiplet is dynamical, while the second
one is semi-dynamical.  It turns out that the corresponding Lagrangians are formulated most directly in the standard ${\cal N}=4$ superspace and its chiral
and anti-chiral subspaces, without applying to the harmonic formalism.  This system is considered here for the first time and it can be regarded as the main new result of our paper.

\subsection{Dynamical mirror multiplet (2, 4, 2)}
The chiral ${\cal N}=4$ superfield as a solution of the constraints \eqref{chiral} is  written as
\bea
    Z\left(t_{\rm L},\theta_i\right)=z+\sqrt{2}\,\theta_{k}\xi^k+\theta_{k}\theta^k B.\label{Z242}
\eea
The relevant off-shell supersymmetry transformations are
\bea
    \delta z = -\,\sqrt{2}\,\epsilon_k\xi^{k},\quad
    \delta \xi^{i} =  \sqrt{2}\,i\bar{\epsilon}^i\dot{z}-\sqrt{2}\,\epsilon^i B,\quad
    \delta B = -\,\sqrt{2}\,i\bar{\epsilon}_k\dot{\xi}^{k}.\label{offsh}
\eea
The total action for the multiplet ${\bf (2, 4, 2)}$ can involve the kinetic and superpotential parts:
\bea
    S_{\bf (2,4,2)}=\frac{1}{4}\int dt\,d\bar{\theta}^2\,d\theta^2\,K\left(Z,\bar{Z}\right)
    +\frac{1}{2}\int dt_{\rm L}\,d^2\theta\,{\cal K}\left(Z\right)+\frac{1}{2}\int dt_{\rm R}\,d^2\bar{\theta}\,\bar{\cal K}\left(\bar{Z}\right).
\eea
The corresponding off-shell component Lagrangian reads
\bea
    {\cal L}_{\bf (2,4,2)} &=& g\left[\dot{\bar{z}}\dot{z} + \frac{i}{2}\left(\xi^{k}\dot{\bar{\xi}}_{k}
    - \dot{\xi}^{k}\bar{\xi}_{k}\right)+ \bar{B}B\right] + \frac{i}{2}\left(\dot{\bar z}\,\partial_{\bar z}g
    - \dot{z}\,\partial_{z} g\right)\xi^{k}\bar{\xi}_{k}+ \frac{\bar{B}}{2}\,\partial_{z}g\,\xi^{k}\xi_k
    + \frac{B}{2}\,\partial_{\bar z}g\,\bar{\xi}_{k}\bar{\xi}^k\nn
    &&+\,\frac{1}{4}\,\partial_{z}\partial_{\bar z}g\,\xi^{i}\xi_i\,\bar{\xi}_{j}\bar{\xi}^j+
    \bar{B}\,\partial_{\bar z}\bar{\cal K}+B\,\partial_{z}{\cal K}-\frac{1}{2}\,\xi_{k}\xi^{k}\,\partial_{z}\partial_{z}{\cal K}
    -\frac{1}{2}\,\bar{\xi}^{k}\bar{\xi}_{k}\,\partial_{\bar z}\partial_{\bar z}{\cal K}\,,\label{242L}
\eea
where $g:=g\left(z,\bar{z}\right)=\partial_{z}\partial_{\bar z}K\left(z,\bar{z}\right)$.

\subsection{Spin mirror multiplet (3, 4, 1) in the chiral superspace}
The triplet superfield $V^{\alpha\beta}$  defined in \eqref{341} can be split into complex and real superfields as
\bea
    V_{12}=Y,\qquad V_{11}=-\,\sqrt{2}\,U,\qquad V_{22}=\sqrt{2}\,\bar{U}.\label{1+2SF}
\eea
The constraints \eqref{341} are rewritten as
\bea
    D^i\bar{U}=0,\qquad \bar{D}_i U=0,\qquad\sqrt{2}\,D_i Y =\bar{D}_i\bar{U},\qquad \sqrt{2}\,\bar{D}_i Y =-\,D_i U,
\eea
where $D^i=D^{i1}$, $\bar{D}^i=D^{i2}$. 
Passing to the new basis at the component level
\bea
    &&v_{12}=y,\qquad v_{11}=-\,\sqrt{2}\,u,\qquad v_{22}=\sqrt{2}\,\bar{u},\nn
    &&\chi^i_1=-\,2\chi^i,\qquad \chi_{j2}=2\bar{\chi}_j\,,
    \qquad C=C,\nn
    &&\epsilon_{i}:=\epsilon_{i1}\,,\qquad \bar{\epsilon}^i=\epsilon^{i}_{2}\,.\label{yububasis}
\eea
we rewrite the off-shell transformations \eqref{341} as 
\bea
    &&\delta u = -\,\sqrt{2}\,\epsilon_k\chi^k,\qquad
    \delta \bar{u} = \sqrt{2}\,\bar{\epsilon}^k\bar{\chi}_k\,,\qquad
    \delta y = \bar{\epsilon}_k\chi^k + \epsilon^k\bar{\chi}_k\,,\nn
    &&\delta \chi^i = \sqrt{2}\,i\bar{\epsilon}^i\dot{u} + \frac{\epsilon^i}{2}\left(C+2i\dot{y}\right),\qquad
    \delta \bar{\chi}_j = -\,\sqrt{2}\,i\epsilon_j\dot{\bar{u}}-\frac{\bar{\epsilon}_j}{2}\left(C-2i\dot{y}\right),\nn
    &&\delta C = 2i\left(\bar{\epsilon}_k\dot{\chi}^k-\epsilon^k\dot{\bar{\chi}}_k\right).\label{yubutr}
\eea
Obviously the complex superfield $U$ is chiral:
\bea
    U\left(t_{\rm L},\theta_i\right)=u+\sqrt{2}\,\theta_{k}\chi^k -\frac{1}{2\sqrt{2}}\,\theta_{k}\theta^k\left(C+2i\dot{y}\right).
\eea
Since the superfield $U$ is chiral we can construct a superpotential as a real sum of the integrals over chiral and anti-chiral subspaces of the ${\cal N}=4, d=1$ superspace:
\bea
    S_{\bf pot.}=\int dt_{\rm L}\,d^2\theta\,{\cal M}\left(U\right)+\int dt_{\rm R}\,d^2\bar{\theta}\,\bar{\cal M}\left(\bar{U}\right).\label{potU}
\eea
It results in a WZ type Lagrangian
\bea
    {\cal L}_{\rm pot.}=-\left[\sqrt{2}\,i\dot{y}\left(\partial_{u}{\cal M}-\partial_{\bar u}\bar{\cal M}\right)
    +\frac{C}{\sqrt{2}}\left(\partial_{u}{\cal M}+\partial_{\bar u}\bar{\cal M}\right)+\chi_{k}\chi^{k}\,\partial_{u}\partial_{u}{\cal M}
    +\bar{\chi}^{k}\bar{\chi}_{k}\,\partial_{\bar u}\partial_{\bar u}{\cal M}\right],\label{WZL-uu}
\eea
that in fact coincides with a particular choice of \eqref{WZL}, with
\bea
    {\cal U}\left(u,\bar{u}\right)=-\,\frac{1}{\sqrt{2}}\left[\partial_{u}{\cal M}\left(u\right)+\partial_{\bar u}\bar{\cal M}\left(\bar{u}\right)\right].
\eea
Thus the superpotential term can be ignored, since it is already present in  \eqref{WZL}.

\subsection{Interaction}
The interaction term for chiral superfields is also written as a superpotential:
\bea
    S_{\rm int.}=\frac{\mu}{2}\int dt_{\rm L}\,d^2\theta\,{\cal F}\left(Z,U\right)+\frac{\mu}{2}\int dt_{\rm R}\,d^2\bar{\theta}\,\bar{\cal F}
    \left(\bar{Z},\bar{U}\right).\label{242int}
\eea
The component Lagrangian reads
\bea
    {\cal L}_{\rm int.}&=&\mu\,\bigg[\,\bar{B}\,\partial_{\bar z}\bar{\cal F}+B\,\partial_{z}{\cal F}
    -\frac{i\dot{y}}{\sqrt{2}}\left(\partial_{u}{\cal F}-\partial_{\bar u}\bar{\cal F}\right)-\frac{C}{2\sqrt{2}}\left(\partial_{u}{\cal F}
    +\partial_{\bar u}\bar{\cal F}\right)\nn
    &&-\,\chi_{k}\xi^{k}\,\partial_{u}\partial_{z}{\cal F}-\frac{1}{2}\,\xi_{k}\xi^{k}\,\partial_{z}\partial_{z}{\cal F}
    -\frac{1}{2}\,\chi_{k}\chi^{k}\,\partial_{u}\partial_{u}{\cal F}\nn
    &&-\,\bar{\chi}^{k}\bar{\xi}_{k}\,\partial_{\bar u}\partial_{\bar z}{\cal F}-\frac{1}{2}\,\bar{\xi}^{k}\bar{\xi}_{k}\,\partial_{\bar z}\partial_{\bar z}{\cal F}
    -\frac{1}{2}\,\bar{\chi}^{k}\bar{\chi}_{k}\,\partial_{\bar u}\partial_{\bar u}{\cal F}\,\bigg]\,.\label{intUZ}
\eea
Note that the interaction Lagrangian ${\cal L}_{\rm int.}$ contains a term $\sim\dot{y}$, {\it i.e.} it can be formally called the interacting WZ Lagrangian.

The total Lagrangian is a sum of \eqref{WZL}, \eqref{242L} and \eqref{intUZ}:
\bea
    {\cal L}_{\rm total}={\cal L}_{\bf (2,4,2)}+{\cal L}_{\rm WZ}+{\cal L}_{\rm int.}\,.
\eea
The function ${\cal F}\left(z,u\right)$ can start with the holomorphic parts ${\cal F}_1\left(z\right)$ and ${\cal F}_2\left(u\right)$. However, their contributions
are identical to those from the corresponding parts of \eqref{242L} and \eqref{WZL-uu}. So they have been already accounted for by ${\cal L}_{\bf (2,4,2)}$ and \eqref{WZL}.
Keeping this in mind, we assume that such parts are absent in the interaction Lagrangian.

For simplicity, when passing to the Hamiltonian formulation, we will limit our consideration to the bosonic constraints:
\bea
    &&\pi_{u}=p_{u}+\sqrt{2}\,i\,{\cal A}_u\approx 0,\qquad
    \pi_{\bar u}=p_{\bar u}-\sqrt{2}\,i\,{\cal A}_{\bar u} \approx 0,\nn
    &&\pi_{y}=p_{y}-i\,{\cal A}_y+\frac{i\mu}{\sqrt{2}}\,\left[\partial_{u}{\cal F}\left(z,u\right)
    -\partial_{\bar u}\bar{\cal F}\left(\bar{z},\bar{u}\right)\right]\approx 0,\nn
    &&h={\cal U}\left(y,u,\bar{u}\right)-\frac{\mu}{2\sqrt{2}}\left[\partial_{u}{\cal F}\left(z,u\right)
    +\partial_{\bar u}\bar{\cal F}\left(\bar{z},\bar{u}\right)\right] \approx 0.\label{hUFbF0}
\eea
Here the last constraint imposes a more complicated relation between the dynamical complex boson $z$ and the semi-dynamical triplet $\left(y,u,\bar{u}\right)$.

The matrix of the constraints \eqref{hUFbF0} is defined as
\bea
\begin{pmatrix}
0& \left\lbrace \pi_{u},\pi_{\bar u} \right\rbrace_{\rm PB}&\left\lbrace \pi_{u},\pi_{y} \right\rbrace_{\rm PB} & \left\lbrace \pi_{u},h \right\rbrace_{\rm PB}\\
\left\lbrace \pi_{\bar u},\pi_{u} \right\rbrace_{\rm PB}&0 &\left\lbrace \pi_{\bar u},\pi_{y} \right\rbrace_{\rm PB} & \left\lbrace \pi_{\bar u},h \right\rbrace_{\rm PB}\\
\left\lbrace \pi_{y},\pi_{u} \right\rbrace_{\rm PB}&\left\lbrace \pi_{y},\pi_{\bar u} \right\rbrace_{\rm PB} & 0 & \left\lbrace \pi_{y},h \right\rbrace_{\rm PB}
\\
\left\lbrace h,\pi_{u} \right\rbrace_{\rm PB}&\left\lbrace h,\pi_{\bar u} \right\rbrace_{\rm PB} & \left\lbrace h,\pi_{y} \right\rbrace_{\rm PB} & 0
\end{pmatrix}
.\label{uuyh}
\eea
Calculating its inverse (see Appendix \ref{AppC}), we obtain the following Dirac brackets
\bea
    &&\left\lbrace z, p_z\right\rbrace = 1,\qquad\left\lbrace p_{z}, y\right\rbrace
    =-\,\frac{\mu\,\partial_{u}\partial_{z}{\cal F}\,\partial_y{\cal U}}{2\sqrt{2}\left(\partial{\cal U}\right)^{2}}\,,
    \qquad\left\lbrace p_{z}, u\right\rbrace =-\,\frac{\mu\,\partial_{u}\partial_{z}{\cal F}}{\sqrt{2}
    \left(\partial{\cal U}\right)^{2}}\left(\partial_{\bar{u}}{\cal U}-\frac{\mu\,\partial_{\bar u}\partial_{\bar u}\bar{\cal F}}{2\sqrt{2}}\right),\nn
    &&\left\lbrace {\bar z}, p_{\bar z}\right\rbrace = 1,\qquad\left\lbrace p_{\bar z}, y\right\rbrace
    =-\,\frac{\mu\,\partial_{\bar u}\partial_{\bar z}\bar{\cal F}\,\partial_y{\cal U}}{2\sqrt{2}\left(\partial{\cal U}\right)^{2}}\,,
    \qquad \left\lbrace p_{\bar z}, \bar{u}\right\rbrace =-\,\frac{\mu\,\partial_{\bar u}\partial_{\bar z}\bar{\cal F}}{\sqrt{2}
    \left(\partial{\cal U}\right)^{2}}\left(\partial_u{\cal U}-\frac{\mu\,\partial_{u}\partial_{u}{\cal F}}{2\sqrt{2}}\right),\nn
    &&\left\lbrace p_z, p_{\bar z}\right\rbrace = -\,\frac{i\mu^2\,\partial_{u}\partial_{z}{\cal F}\,
    \partial_{\bar u}\partial_{\bar z}\bar{\cal F}\,\partial_y{\cal U}}{2\left(\partial{\cal U}\right)^{2}}\,,\qquad\left\lbrace u, \bar{u}\right\rbrace
    = -\,\frac{i\,\partial_y{\cal U}}{2\left(\partial{\cal U}\right)^{2}}\,,\nn
    &&\left\lbrace y, u\right\rbrace = -\,\frac{i}{2\left(\partial{\cal U}\right)^{2}}\left(\partial_{\bar{u}}{\cal U}
    -\frac{\mu\,\partial_{\bar u}\partial_{\bar u}\bar{\cal F}}{2\sqrt{2}}\right),\qquad \left\lbrace y, \bar{u}\right\rbrace =
    \frac{i}{2\left(\partial{\cal U}\right)^{2}}\left(\partial_u{\cal U}-\frac{\mu\,\partial_{u}\partial_{u}{\cal F}}{2\sqrt{2}}\right),\nn
    &&\left(\partial{\cal U}\right)^{2}=\left[\partial_y{\cal U}\,\partial_y{\cal U}+2\left(\partial_{\bar{u}}{\cal U}
    -\frac{\mu\,\partial_{\bar u}\partial_{\bar u}\bar{\cal F}}{2\sqrt{2}}\right)\left(\partial_u{\cal U}-\frac{\mu\,\partial_{u}\partial_{u}{\cal F}}{2\sqrt{2}}\right)\right].\label{Brack1}
\eea
One can make use of the identity
\bea
   \left\lbrace p_{z}, y\right\rbrace\, \partial_{\bar u}\partial_{\bar z}\bar{\cal F}
    = \left\lbrace p_{\bar z}, y\right\rbrace\,\partial_{u}\partial_{z}{\cal F}
\eea
in order to simplify \eqref{Brack1} to the form
\bea
    &&\left\lbrace z, p_z\right\rbrace = 1,\qquad\left\lbrace {\bar z}, p_{\bar z}\right\rbrace = 1,\qquad\left\lbrace p_z, p_{\bar z}\right\rbrace
    = \frac{i}{\sqrt{2}}\,\mu\left(\left\lbrace p_{z}, y\right\rbrace\partial_{\bar u}\partial_{\bar z}\bar{\cal F}
    +\left\lbrace p_{\bar z}, y\right\rbrace\partial_{u}\partial_{z}{\cal F}\right),\nn
    &&\left\lbrace p_{z}, y\right\rbrace = -\,\frac{i}{\sqrt{2}}\,\mu\,\left\lbrace u,\bar{u}\right\rbrace\,\partial_{u}\partial_{z}{\cal F},\qquad
    \left\lbrace p_{\bar z}, y\right\rbrace = -\,\frac{i}{\sqrt{2}}\,\mu\,\left\lbrace u,\bar{u}\right\rbrace\,\partial_{\bar u}\partial_{\bar z}\bar{\cal F} ,\nn
    &&\left\lbrace p_{z}, u\right\rbrace = -\,\sqrt{2}\,i\mu\,\left\lbrace y,\bar{u}\right\rbrace\,\partial_{u}\partial_{z}{\cal F}\,,\qquad
     \left\lbrace p_{\bar z}, \bar{u}\right\rbrace = \sqrt{2}\,i\mu\,\left\lbrace y,\bar{u}\right\rbrace\,\partial_{\bar u}\partial_{\bar z}\bar{\cal F}\,.\label{Brack2}
\eea
These are a generalization of the Nahm equations \eqref{NahmEqs} with a complex evolution parameter $z$.

The simplest non-commutative plane solution \eqref{noncbr} requires that
\bea
    \partial_u{\cal U}\left(y,u,\bar{u}\right)-\frac{\mu}{2\sqrt{2}}\,\partial_{u}\partial_{u}{\cal F}\left(u,z\right)=0,\qquad \partial_{\bar{u}}{\cal U}\left(y,u,\bar{u}\right)-\frac{\mu}{2\sqrt{2}}\,\partial_{\bar u}\partial_{\bar u}\bar{\cal F}\left(\bar{u},\bar{z}\right)=0\quad
\Rightarrow\nn
\Rightarrow\quad
    \partial_{u}{\cal U}=0,\quad \partial_{\bar u}{\cal U}=0,\quad\partial_{u}\partial_{u}{\cal F}=0,\quad \partial_{\bar u}\partial_{\bar u}\bar{\cal F}=0.
\eea
It yields \eqref{nonc} and fixes the function ${\cal F}$ as
\bea
    {\cal F}\left(z,u\right)=u\,{\cal S}\left(z\right),\quad\bar{\cal F}\left(\bar{z},\bar{u}\right)=\bar{u}\,\bar{\cal S}\left(\bar{z}\right).
\eea
So we obtain
\bea
    \left\lbrace z, p_z\right\rbrace = 1,\qquad
    \left\lbrace {\bar z}, p_{\bar z}\right\rbrace = 1,\qquad
    \left\lbrace p_z, p_{\bar z}\right\rbrace = 2i\mu^2\,\partial_{z}{\cal S}\,\partial_{\bar z}\bar{\cal S},\qquad
    \left\lbrace u, \bar{u}\right\rbrace = i.
\eea
The third spin variable $y$ is now represented as a function of the dynamical boson $z$:
\bea
    y=c-\frac{\mu}{\sqrt{2}}\left[{\cal S}\left(z\right)
    +\bar{\cal S}\left(\bar{z}\right)\right].
\eea

\section{Summary and outlook}
In this paper we elucidated the distinction between ordinary and mirror multiplets of ${\cal N}=4$, $d=1$ supersymmetry. Mirror multiplets
are described by superfields carrying no external SU$(2)_{\rm L}$ indices and satisfying the common constraint \eqref{mirror} as a consequence of their basic constraints
linear in the covariant spinor derivatives. According to this general definition the standard chiral multiplet ${\bf (2, 4, 2)}$ belongs to the mirror type.

We considered the mirror multiplet ${\bf (3, 4, 1)}$ as semi-dynamical and constructed its action \eqref{WZ-SF} in the analytic
harmonic superspace as a particular case of the general WZ action for mirror multiplets \eqref{WZ}. We coupled it to the dynamical mirror multiplet ${\bf (1, 4, 3)}$ and constructed
their interaction \eqref{int-SF} in the analytic harmonic superspace. We obtained a mirror analogue of the model studied in \cite{FIL2012} and considered
its deformation to ${\rm SU}(2|1)$ supersymmetry.

We constructed, for the first time,  the coupling of the semi-dynamical mirror multiplet ${\bf (3, 4, 1)}$ to the dynamical mirror multiplet ${\bf (2, 4, 2)}$ in the chiral
${\cal N}=4, d=1$ superspace.
We calculated Dirac brackets for the spin variables, dynamical fields and their momenta. These brackets accomplish a kind of generalization of Nahm equations associated
with the previously considered ${\bf (3, 4, 1)} - {\bf (1, 4, 3)}$ system, such that the complex $d=1$ field $z$ plays now the role of complex evolution parameter.
 It would be interesting to study this model in more detail, in particular to calculate its supercharges, and to pass to its quantum version. Also, the possible deformed ${\rm SU}(2|1)$ version of this new system seems to deserve
 an attention \footnote{${\rm SU}(2|1)$ invariant system  ${\bf (4, 4, 0)} - {\bf (2, 4, 2)}$, with both multiplets being mirror, was constructed in \cite{SidSolo}.}.

The  most intriguing question for further study is whether it is possible to construct ${\cal N}=4$ SQM models involving interactions between the multiplets which are mirror
to each other. Perhaps, the bi-harmonic formalism of ref. \cite{BHSS} augmented with the consideration in the present paper could help to advance towards this
goal \footnote{To date, only one example of such a system with non-trivial self-interaction is known \cite{BHSS}.}.

\section*{Acknowledgements}
The authors thank Sergey Fedoruk for useful discussions and comments. The research
was supported by the Russian Science Foundation Grant No 21-12-00129.
Stepan Sidorov thanks Yerevan Physics Institute for a kind hospitality at the final stage of completing the paper.

\appendix
\section{Basics of ${\cal N}=4$, $d=1$ superspace}\label{AppA}
The coordinates of ${\cal N}=4$, $d=1$ superspace $\zeta :=\left\lbrace t, \theta^{i\alpha}\right\rbrace$ transform under ${\cal N}=4$ supersymmetry  as
\bea
    \delta \theta^{i\alpha}= \epsilon^{i\alpha}, \qquad  \delta t = -\,i\epsilon^{i\alpha}\theta_{i\alpha}\,,
    \qquad \overline{\left(\theta^{i\alpha}\right)}=-\,\theta_{i\alpha}\,,\qquad \overline{\left(\epsilon^{i\alpha}\right)}
    =-\,\epsilon_{i\alpha}\,, \label{SStr}
\eea
where $\epsilon^{i\alpha}$ is a quartet of the corresponding Grassmann parameters. The covariant derivatives are defined as
\bea
    D^{i\alpha}=\frac{\partial}{\partial \theta_{i\alpha}}+i\theta^{i\alpha}\partial_t\,.
\eea

The coordinates of the left-handed chiral subspace $\zeta_{\rm L} :=\left\lbrace t_{\rm L}, \theta_{i}\right\rbrace$ are related to the previously defined ones as
\bea
    t_{\rm L}:=t-i\,\theta^{i1}\theta_{i1},\qquad\theta_{i}:=\theta_{i1}\,.
\eea
They transform as
\bea
    \delta \theta_{i}= \epsilon_{i}\,, \qquad  \delta t_{\rm L} = 2i\bar{\epsilon}^{k}\theta_{k}\,,\qquad \overline{\left(\epsilon_{i}\right)}=\bar{\epsilon}^{i}.
\eea
\subsection{Harmonic superspace}
We perform harmonization over the indices corresponding to ${\rm SU}(2)_{\rm L}$\,:
\bea
    t_{\rm (A)}=t-\frac{i}{2}\,\theta^{i}_{\alpha}\theta^{j\alpha}
    \left(u^+_iu^-_j + u^+_j u^-_i\right),\qquad \theta^{\pm}_{\alpha} := \theta^{i}_{\alpha}u^{\pm}_{i},\qquad
    u^+_iu^-_j - u^+_j u^-_i = \varepsilon_{ij}\,.
\eea
Then the harmonic superspace is defined by
\bea
    \zeta_{\rm H} :=\left\lbrace t_{\rm (A)}, \theta^{\pm}_{\alpha}, u^{\pm}_i\right\rbrace.\label{HSS}
\eea
Its coordinates transform as
\bea
    \delta \theta^{\pm}_{\alpha}= \epsilon^{\pm}_{\alpha}, \qquad
    \delta u^{\pm}_{i}=0,\qquad  \delta t_{\rm (A)}
    = 2i\epsilon^{-\alpha}\theta^{+}_{\alpha},\qquad \epsilon^{\pm}_{\alpha} := \epsilon^{i}_{\alpha}u^{\pm}_{i}\,. \label{HSStr}
\eea

The measure of integration over the full harmonic superspace is defined as
\bea
    d\zeta_{\rm H} := \frac{1}{4}\,du\, dt_{\rm (A)} \,d\theta^{+}_{\alpha}d\theta^{+\alpha}d\theta^{-}_{\beta}d\theta^{-\beta}.
\eea
The harmonic superspace involves the analytic harmonic subspace parametrized by the reduced coordinate set
\bea
    \zeta_{\rm (A)} :=\left\lbrace t_{\rm (A)},\theta^{+}_{\alpha}, u^{\pm}_i\right\rbrace,
\eea
which is closed under the transformations \eqref{HSStr}.

We use the standard notation for covariant derivatives
\bea
    &&D^{+\alpha} = \frac{\partial}{\partial \theta^{-}_{\alpha}}\,,\nn
    &&D^{++} = \partial^{++} - i\,\theta^{+}_{\alpha}\theta^{+\alpha}\frac{\partial}{\partial t_{\rm (A)}}
    + \theta^{+}_{\alpha}\frac{\partial}{\partial \theta^{-}_\alpha} \,,\nn
    &&D^0 = \partial^0 + \theta^{+}_{\alpha}\frac{\partial}{\partial \theta^{+}_{\alpha}}
- \theta^{-}_{\alpha}\frac{\partial}{\partial \theta^{-}_{\alpha}}\,,
\eea
where the partial harmonic derivatives are
\bea
&&    \partial^{\pm\pm} := u^\pm_i \frac{\partial}{\partial u^\mp_i} \,,
    \qquad  \partial^0 := u^+_i \frac{\partial}{\partial u^+_i} - u^-_i \frac{\partial}{\partial u^-_i}\,,\nn
&&    [\partial^{++}, \partial^{--}] = \partial^0, \quad [\partial^0, \partial^{\pm\pm}] = \pm\,2 \partial^{\pm\pm}\,.
\eea
According to these definitions, the invariant integration measure of the analytic subspace $d\zeta^{--}_{\rm (A)}$ is related to $d\zeta_{\rm H}$ as
\bea
    d\zeta^{--}_{\rm (A)} := \frac{1}{2}\,du\, dt_{\rm (A)} \,d\theta^{+}_{\alpha}d\theta^{+\alpha},\qquad
    d\zeta_{\rm H} = \frac{1}{2}\,d\zeta^{--}_{\rm (A)}\,D^{+}_{\alpha}D^{+\alpha}.
\eea

\section{Component solutions for the mirror multiplets (4, 4, 0) and (0, 4, 4)}\label{AppB}
\paragraph{Mirror multiplet ${\bf (4,4,0)}$.} The constraints \eqref{440HSS} are solved by
\bea
    Y^{\alpha A}=y^{\alpha A}+\theta^{-\alpha}\psi^{iA}u^{+}_i-\theta^{+\alpha}\psi^{iA}u^{-}_i+2i\,\theta^{+}_{\beta}\theta^{-\alpha}\dot{y}^{\beta A}+i\,\theta^{+}_{\beta}\theta^{+\beta}\theta^{-\alpha}\dot{\psi}^{iA}u^{-}_i.
\eea
The components transform as
\bea
	\delta y^{\alpha A} = -\,\epsilon^{\alpha}_{k}\psi^{kA},\qquad
	\delta \psi^{iA} = 2i\epsilon^{i\gamma}\dot{y}^{A}_{\gamma}.
\eea
The corresponding SU$(2|1)$ deformed solution and the transformation properties are given in \cite{DHSS}.

\paragraph{Mirror multiplet ${\bf (0,4,4)}$.} The solution of eq. \eqref{044} for the fermionic superfield $\Psi^{iA}$ is written as
\bea
    \Psi^{\alpha A}=\psi^{\alpha A}+\theta^{-\alpha}D^{iA}u^{+}_i-\theta^{+\alpha}D^{iA}u^{-}_i+2i\,\theta^{+}_{\beta}\theta^{-\alpha}\dot{\psi}^{\beta A}+i\,\theta^{+}_{\beta}\theta^{+\beta}\theta^{-\alpha}\dot{D}^{iA}u^{-}_i.
\eea
The transformation properties of the component fields are
\bea
	\delta \psi^{\alpha A} = -\,\epsilon^{\alpha}_{k}D^{kA},\qquad
	\delta D^{iA} = 2i\epsilon^{i\gamma}\dot{\psi}^{A}_{\gamma}.
\eea

\section{Matrices of second-class constraints}\label{AppC}
The matrix \eqref{uuyh} in the explicit form reads:
\bea
\begin{pmatrix}
0 & -\,2i\partial_y{\cal U} & 2i\partial_u{\cal U}-\frac{i\mu\,\partial_{u}\partial_{u}{\cal F}}{\sqrt{2}}
& -\,\partial_u{\cal U}+\frac{\mu\,\partial_{u}\partial_{u}{\cal F}}{2\sqrt{2}}\\
2i\partial_y{\cal U} & 0 & -\,2i\partial_{\bar{u}}{\cal U}+\frac{i\mu\,\partial_{\bar u}\partial_{\bar u}\bar{\cal F}}{\sqrt{2}}
& -\,\partial_{\bar{u}}{\cal U}+\frac{\mu\,\partial_{\bar u}\partial_{\bar u}\bar{\cal F}}{2\sqrt{2}}\\
-\,2i\partial_u{\cal U}+\frac{i\mu\,\partial_{u}\partial_{u}{\cal F}}{\sqrt{2}} & 2i\partial_{\bar{u}}{\cal U}
-\frac{i\mu\,\partial_{\bar u}\partial_{\bar u}\bar{\cal F}}{\sqrt{2}} & 0 & -\,\partial_y{\cal U}\\
\partial_u{\cal U}-\frac{\mu\,\partial_{u}\partial_{u}{\cal F}}{2\sqrt{2}} & \partial_{\bar{u}}{\cal U}
-\frac{\mu\,\partial_{\bar u}\partial_{\bar u}\bar{\cal F}}{2\sqrt{2}}  & \partial_y{\cal U} & 0\\
\end{pmatrix}.\label{matrix}
\eea
The corresponding  inverse matrix is then calculated  to be
\bea
\frac{1}{\left(\partial{\cal U}\right)^{2}}
\begin{pmatrix}
0 & -\,\frac{i}{2}\,\partial_y{\cal U} & \frac{i}{2}\,\partial_{\bar{u}}{\cal U}
-\frac{i\mu\,\partial_{\bar u}\partial_{\bar u}\bar{\cal F}}{4\sqrt{2}} & \partial_{\bar{u}}{\cal U}
-\frac{\mu\,\partial_{\bar u}\partial_{\bar u}\bar{\cal F}}{2\sqrt{2}}\\
\frac{i}{2}\,\partial_y{\cal U} & 0 & -\,\frac{i}{2}\,\partial_u{\cal U}+\frac{i\mu\,\partial_{u}\partial_{u}{\cal F}}{4\sqrt{2}}
& \partial_u{\cal U}-\frac{\mu\,\partial_{u}\partial_{u}{\cal F}}{2\sqrt{2}}\\
-\,\frac{i}{2}\,\partial_{\bar{u}}{\cal U}+\frac{i\mu\,\partial_{\bar u}\partial_{\bar u}\bar{\cal F}}{4\sqrt{2}} & \frac{i}{2}\,\partial_u{\cal U}
-\frac{i\mu\,\partial_{u}\partial_{u}{\cal F}}{4\sqrt{2}} & 0 & \partial_y{\cal U}\\
-\,\partial_{\bar{u}}{\cal U}+\frac{\mu\,\partial_{\bar u}\partial_{\bar u}\bar{\cal F}}{2\sqrt{2}}
& -\,\partial_u{\cal U}+\frac{\mu\,\partial_{u}\partial_{u}{\cal F}}{2\sqrt{2}}& -\,\partial_y{\cal U} & 0\\
\end{pmatrix}.\label{inverse}
\eea

\end{document}